\documentclass[osajnl,preprint,showpacs]{revtex4}
\usepackage{graphicx}

\begin{document}

\title{Coupled-mode theory for spatial gap solitons in optically-induced
lattices}

\author{Boris A. Malomed}

\affiliation{Department of Interdisciplinary Studies, School of
Electrical Engineering, Faculty of Engineering, Tel Aviv
University, Tel Aviv 69978, Israel}

\author{Elena A. Ostrovskaya and Yuri S. Kivshar}

\affiliation{Nonlinear Physics Center, Research School of Physical Sciences and
Engineering, Australian National University, Canberra ACT 0200, Australia}

\author{T. Mayteevarunyoo}

\affiliation{Department of Telecommunication Engineering,
Mahanakorn University of Technology, Bangkok 10530, Thailand}

\begin{abstract}
We develop a coupled-mode theory for spatial gap solitons in the
one-dimensional photonic lattices induced by interfering optical
beams in a nonlinear photorefractive crystal. We derive a novel
system of coupled-mode equations for two counter-propagating probe
waves, and find its analytical solutions for stationary gap
solitons. We also predict the existence of moving (or tilted) gap
solitons and study numerically soliton collisions.
\end{abstract}

\maketitle

%\ocis{030.1640, 190.4420}

It is well known that a periodic modulation of the optical
refractive index not only modifies the spectrum of linear waves,
but also strongly affects nonlinear propagation and self-trapping
of light~\cite{book0,book}. Recently, formation of spatial
solitons in reconfigurable photonic lattices in photorefractive
materials has been demonstrated experimentally in
one-~\cite{Fleischer:2003-23902:PRL,Neshev:2003-710:OL} and two-
\cite{Fleischer:2003-147:Nature} dimensional geometries. In this
case, strong electro-optic anisotropy of a photorefractive crystal
was employed to create the lattice by interfering laser beams with
ordinary polarization, while the solitons were observed in the
orthogonally polarized mode.

Periodically modulated nonlinear systems can also support
self-trapped localized pulses or beams in the form of \emph{gap
solitons} (GSs), which are hosted by a gap of the system's linear
spectrum, induced by the resonant Bragg coupling between the
forward- and backward-propagating waves~\cite{deSterke}. A notable
property of the GSs are that they can form in both self-focusing
and self-defocusing media.

Travelling temporal-domain GSs have been observed experimentally
in material Bragg gratings written in silica
fibers~\cite{Eggleton:1996-1627:PRL}. The concept of a
spatial-domain GS was proposed~\cite{Feng,Nabiev} and further
elaborated in waveguide
settings~\cite{kivshar,Mak-spatial,sukhorukov}. Experimentally,
spatial GSs were demonstrated in waveguide arrays~\cite{Mandelik}
and optically-induced photonic lattices~\cite{gap_induced}.

The simplest and most ubiquitous description of the GSs is
provided by the coupled-mode theory (CMT), which amounts to the
derivation of a system of coupled nonlinear propagation equations
for the forward and backward waves~\cite{deSterke}. In this
Letter, we derive a novel type of the CMT model for spatial GSs in
the optically-induced photonic lattice. We find {\em exact
analytical solutions} for GSs in the framework of this model, and
compare them with the results obtained numerically by solving the
full nonlinear model. We thus identify a parameter region where
the couple-mode approximation is accurate enough, and in that
region we use the model to predict new features such the existence
of the families of moving (or tilted) gap solitons and to study
their interaction Actually, the new model derived in this work may
have more general purport than just an asymptotic approximation
for the nonlinear photonic lattice, as it is the first
coupled-mode system that accounts for a saturable nonlinearity.

Following
Refs.~\cite{Fleischer:2003-23902:PRL,Neshev:2003-710:OL}, we
consider the evolution of an extraordinarily polarized probe beam
propagating through a periodic structure written by
counter-propagating ordinary-polarized plane-wave beams in the
photorefractive medium. As mentioned above, the electro-optic
coefficients in the crystal are substantially different for the
orthogonal polarizations, therefore the grating, created by
interference of the counter-propagating beams, is essentially
one-dimensional (uniform along the propagation axis $z$). The
interference of the plane waves with a wavelength $\lambda $
creates an intensity distribution, $I_{g}(x)=I_{0}\cos ^{2}(Kx)$,
with $K=2\pi n_{0}\lambda ^{-1}\sin \theta $, where $\theta $ is
the angle between the Poynting vectors of the plane waves and the
$z$ axis, and $n_{0}$ is the refractive index along the ordinary
axis. Provided that the intensity of the probe beam, $|E|^{2}$, is
much weaker than that of the grating, $I_{0}$, one may neglect the
feedback action of the probe beam on the grating (a perturbative
calculation within the framework of an extended model, that
includes equations for both the lattice-forming and probe fields,
shows that the feedback exactly cancels out at the first order in
$1/I_{0}$). Then the evolution of the local amplitude $E(x,z)$ of
the probe beam obeys the known
equation~\cite{Fleischer:2003-23902:PRL,Neshev:2003-710:OL}, whose
normalized form is
\begin{equation}
i\frac{\partial E}{\partial z}+\frac{1}{2}\frac{\partial ^{2}E}{\partial
x^{2}}-\frac{E}{1+I_{0}\cos ^{2}(Kx)+|E|^{2}}=0.
\label{eq1}
\end{equation}

To derive the corresponding equations for the coupled forward and backward
waves, we approximate solutions to Eq.~(\ref{eq1}) by
\begin{equation}
E(x,z)=u(x,z)e^{iKx}+v(x,z)e^{-iKx},
\label{eq2}
\end{equation}
where $u$ and $v$ are slowly varying [in comparison with the
carrier waves $ \exp \left( \pm iKx\right) $] envelopes of the
forward and backward modes. Substituting the expansion (\ref{eq2})
into Eq.~(\ref{eq1}), we perform the Fourier expansion with
respect to $\exp \left( \pm iKx\right)$ and, in the spirit of the
CMT approach, keep only the lowest-order harmonics. Eventually,
this leads to the following CMT equations (one of which is
linear): %%\begin{eqnarray}
%%i\frac{\partial u}{\partial z} &+&iK\frac{\partial u}{\partial x}=\frac{(u-v)%
%%}{\sqrt{I_{0}(1+|u-v|^{2})+1+2\left( |u|^{2}+|v|^{2}\right) }},  \nonumber \\
%%i\frac{\partial v}{\partial z} &-&iK\frac{\partial v}{\partial x}=\frac{(v-u)%
%%}{\sqrt{I_{0}(1+|u-v|^{2})+1+2\left( |u|^{2}+|v|^{2}\right) }}.  \nonumber
%%\end{eqnarray}%
\begin{equation}
i\frac{\partial }{\partial z}(u-v)+iK\frac{\partial }{\partial
x}(u+v) \nonumber
\end{equation}
\begin{equation}
-\frac{2(u-v)}{\sqrt{I_{0}(1+|u-v|^{2})+1+2\left(
|u|^{2}+|v|^{2}\right) }} =0,  \label{nonlinear}
\end{equation}
\begin{equation}
i\frac{\partial }{\partial z}(u+v)+iK\frac{\partial }{\partial x}(u-v)=0.
\label{linear}
\end{equation}
Equations (\ref{nonlinear}) and (\ref{linear}) constitute a
\emph{new CMT model} with the saturable nonlinearity. It contains
one irreducible parameter $I_{0}$, while $K$ can be absorbed by
rescaling of $x$.

In the physically relevant case, the photonic-lattice intensity is
large, i.e., $I_{0}\gg 1,|u|^{2},|v|^{2}$, hence the square root
in Eq. (\ref {nonlinear}) may be approximated by
$\sqrt{I_{0}(1+|u-v|^{2})}$, except for a vicinity of point(s)
where $w\equiv u-v$ vanishes. Using this approximation, and
eliminating $\left( u+v\right) $ by means of Eq. (\ref {linear}),
we reduce Eq. (\ref{nonlinear}) to an equation for the single
function $w$,
\begin{equation}
\frac{\partial ^{2}w}{\partial z^{2}}-K^{2}\frac{\partial
^{2}w}{\partial x^{2}}+\frac{2i}{\sqrt{I_{0}}}\frac{\partial
}{\partial z}\left( \frac{w} {\sqrt{(1+|w|^{2})}}\right) =0.
\label{eq3}
\end{equation}
We have checked the accuracy of the simplified equation
(\ref{eq3}), comparing its analytical solutions for solitons (see
below), and their stability, vs. direct numerical solutions of
Eqs. (\ref{nonlinear}) and (\ref {linear}). A conspicuous
difference appears only in the region of $ I_{0}\,_{\sim
}^{<}\,3$, where the CMT does not provide for an adequate
approximation anyway.

Stationary solutions of Eq.~(\ref{eq3}) are sought for as
$w(x,z)=e^{iqz}W(x) $, where a real function $W$ obeys the
equation $d^{2}W/dx^{2}=-dU_{\mathrm{ eff}}/dW$, with
\begin{equation}
U_{\mathrm{eff}}(W)=\frac{q^{2}}{2K^{2}}W^{2}+\frac{2q}{K^{2}\sqrt{I_{0}}}
\left( \sqrt{1+W^{2}}-1\right) .  \label{eq5}
\end{equation}
Further, Eq. (\ref{linear}) shows that, for the stationary solutions,
\begin{equation}
(u,v)=\frac{1}{2}e^{iqz}\left[ W(x)\pm \frac{iK}{q}\frac{dW}{dx}\right] .
\label{sys}
\end{equation}
If the propagation constant $q$ belongs to the interval (which is, actually,
the bandgap)
\begin{equation}
0<-q<2{\Large /}\sqrt{I_{0}}\equiv Q,  \label{interval}
\end{equation}
the potential (\ref{eq5}) has two symmetric minima, giving rise to GS
solutions that can be written in an implicit analytical form,
\begin{equation}
\left( KW^{\prime }\right) ^{2}=-\left( qW\right) ^{2}+\left(
4|q|{\Large /} \sqrt{I_{0}}\right) \left( \sqrt{1+W^{2}}-1\right)
.  \label{eq6}
\end{equation}

As it follows from Eq. (\ref{eq6}), the soliton's squared
amplitude is
$W_{\mathrm{max}}^{2}=(8/q^{2}I_{0})(2-\sqrt{I_{0}}|q|)$. In the
small-amplitude limit, i.e., for $0<\epsilon \equiv
2(2-|q|\sqrt{I_{0}})\ll 1$ [cf. Eq. (\ref {interval})], the GS
asymptotically coincides with the conventional
nonlinear-Schr\"{o}dinger soliton, $W(x)=\pm \sqrt{\epsilon
}\;\mathrm{sech} \left( \sqrt{\epsilon
|q|/2}K^{-1}I_{0}^{-1/4}x\right) $. In the other limit,
$|q|\rightarrow 0$, the soliton assumes a \textquotedblleft
compacton" shape: $|q|\sqrt{I_{0}}W(x)=4\cos ^{2}\left(
|q|K^{-1}I_{0}^{-1/4}x\right) $, if $|x|<\pi KI_{0}^{1/4}/\left(
2|q|\right) $, and $W(x)=0$ otherwise. However, the conditions
under which the CMT equations were derived above do not hold in
the latter case.

\begin{figure}[tbp]
\includegraphics[width=3.3in]{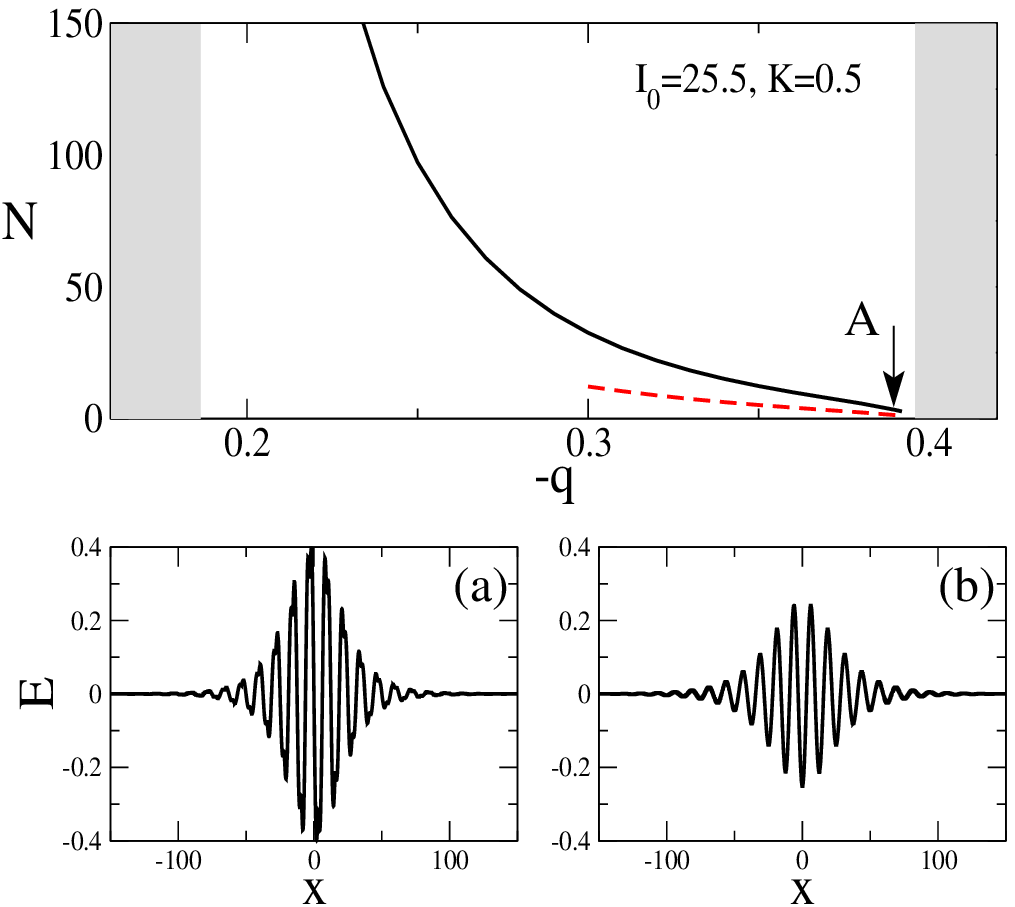}
\caption{The integral power of the soliton vs. the propagation
constant [in the first finite gap of Eq. (\protect\ref{eq1})] at
$I_{0}=25.5$ and $K=0.5$ . The solid and dashed lines show,
respectively, the numerical result, and the analytical one, as
obtained from Eqs. (\protect\ref{eq6}), (\protect\ref {sys}), and
(\protect\ref{eq2}). Examples of numerical (a) and analytical (b)
profiles of a soliton taken close to the edge of the gap
($q=0.39$, point A) are shown in the bottom part of the figure.}
\label{fig1}
\end{figure}

To verify the validity of the CMT approximation in the present
setting, in Fig. \ref{fig1} we compare the analytical soliton
solutions based on Eq. (\ref{eq6}) and ones found numerically from
Eq. (\ref{eq1}). The comparison is presented in terms of a global
characteristic of the soliton family, viz., the integral power,
$N=\int_{-\infty }^{+\infty }|E(x)|^{2}dx$, vs. the propagation
constant $q$; an example of the comparison of the soliton's shapes
is also included. Naturally, the approximation is appropriate
sufficiently close to the gap's edge. We also note that the
negative slope of the $N(q)$ dependence suggests stability of the
entire soliton family as per the Vakhitov-Kolokolov criterion,
i.e., the absence of real eigenvalues in the spectrum of small
perturbations around the soliton \cite{VK}. However, the solitons
may be subject to instabilities with complex eigenvalues. Detailed
numerical analysis shows that, in the case of large $ I_{0}$, when
the the CMT approximation is relevant, a stable part of the
soliton family in the bandgap (\ref{interval}) is limited to a
sub-gap,
\begin{equation}
Q-\Delta Q<-q<Q.  \label{stability}
\end{equation}
The width of the stability sub-gap, $\Delta Q\approx 0.21$, is
nearly constant for $I_{0}\,_{\sim }^{>}\,20$ [for $I_{0}=25.5$,
which is the case shown in Fig. \ref{fig1}, the GSs are stable in
$\approx 53\%$ of the interval (\ref{interval}); we mention, for
comparison, that in the standard GS model with the cubic
nonlinearity, the stable part occupies $\approx 51\%$ of the
bandgap is \cite{Rich,CapeTown}]. Unstable solitons [which
actually occupy that part of the bandgap (\ref{interval}) where
the CMT approximation is irrelevant] are destroyed by
perturbations.

The CMT approximation opens many ways to investigate novel
phenomena which may be difficult to study directly within the
framework of the full model. An issue of obvious interest are
moving (or tilted) GSs, of the form $\left( u,v\right)
=e^{iqz}\left( U(x-cz),V(x-cz)\right) $. Our analysis shows that
the bandgap (\ref{interval}) for the tilted solitons shrinks to
$0<-q<\tilde{Q}\equiv Q \sqrt{1-(c/K)^{2}}$, and it does not exist
for $c^{2}>K^{2}$. The reduced gap is completely filled with
moving gap solitons which are stable in the sub-gap
$\tilde{Q}-\Delta Q<-q<\tilde{Q}$, cf. Eq. (\ref{stability}), with
the width $\Delta Q$ that does not depend on the tilt $c$, up to
the accuracy of numerical results (the latter fact resembles a
known stability feature of the GSs in the conventional CMT
approach~\cite{CapeTown}).

\begin{figure}[tbp]
\includegraphics[width=3.3in]{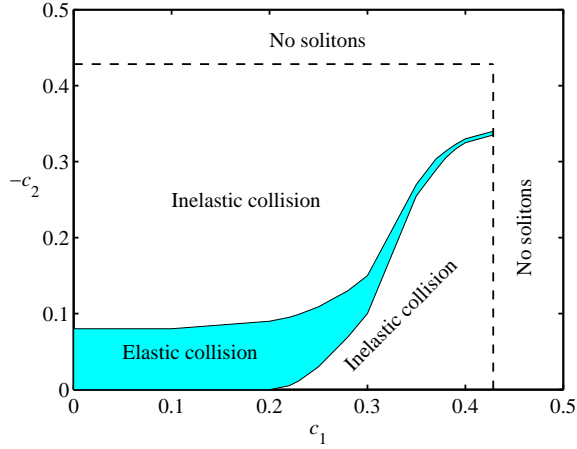}
\caption{Regions of elastic and inelastic head-on collisions
between two solitons, with the tilts $c_{1}>0$ and $c_{2}<0$,  and
$|q_{1}|=|q_{2}|=0.2$ (close to the instability border, which is
$|q|=Q-\Delta Q\approx 0.19$, in this case), at $I_{0}=25.5$ and
$K=0.5$.} \label{fig2}
\end{figure}

The stability of the tilted solitons suggests to consider
collisions (intersections) between them. The most important
characteristic of soliton interactions is elasticity. Our
numerical simulations demonstrate that the interaction is quite
elastic, unless the solitons are taken close to the instability
border [which is $-q=Q-\Delta Q$, see Eq. (\ref{stability})]. In
the latter case, there is a specific structure of regions of
elastic and inelastic collisions in the plane $\left(
c_{1},-c_{2}\right) $ of the tilts of two colliding solitons, as
shown in Fig. \ref{fig2}. In the case of inelasticity, the
colliding solitons suffer strong perturbations.

In conclusion, we have developed the coupled-mode theory for
spatial gap solitons in one-dimensional photonic lattices written
optically in a photorefractive medium. We have derived a novel
type of the coupled-mode equations and found their analytical
solutions for gap solitons. In the case when the coupled-mode
system correctly approximates the photonic lattice, we used the
model to investigate new features of such systems, such as the
existence and stability of moving gap solitons and their
collisions. The latter results suggest new experiments for
optically-induced lattices in photorefractive media.

This work was partially supported by the Australian Research Council. B.A.M.
appreciates hospitality of the Nonlinear Physics Center at the Australian
National University.

\newpage

\newpage

\end{document}